\documentclass[aps,pra,showpacs,twocolumn,9pt]{revtex4-1}
\usepackage{epsfig}
\usepackage{graphics}
\usepackage{graphicx}
\usepackage{epstopdf}
\DeclareGraphicsExtensions{,.pdf,.png,.jpg,.gif}

\begin{document}
\title{Vortices on demand in multicomponent Bose-Einstein condensates}

\author{R. Zamora-Zamora, M. Lozada-Hidalgo, S. F. Caballero-Ben\'itez and V. Romero-Roch\'in} 

\email{romero@fisica.unam.mx}

\affiliation{Instituto de F\'{\i}sica, Universidad
Nacional Aut\'onoma de M\'exico. \\ Apartado Postal 20-364, 01000 M\'exico D.
F. Mexico. }


\date{\today}

\begin{abstract}
We present a simple mechanism to produce vortices at any desired spatial locations in harmonically  trapped Bose-Einstein condensates (BEC) with multicomponent spin states coupled to external transverse and axial magnetic fields. The vortices appear at the spatial points where the spin-transverse field interaction vanishes and, depending on the multipolar magnetic field order, the vortices can acquire different predictable topological charges. We explicitly demonstrate our findings, both numerically and analytically, by analyzing a 2D BEC via the Gross-Pitaevskii equation for atomic systems with either two or three internal states. We further show that, by an spontaneous symmetry breaking mechanism, vortices can appear in any spin component, unless symmetry is externally broken at the outset by an axial field. We suggest that this scenario may be tested using an ultracold gas of $^{87}$Rb occupying all three $F = 1$ states in an optical trap. 

\end{abstract}

\maketitle

\section{Introduction}
\label{sec:Introduction}

Quantum vortices are the hallmark of superfuidity in either fermionic or bosonic ultracold gases\cite{dum,castin,matthews, zwierlein2, cooper, abo, madison}. Following the seminal work of Abrikosov\cite{abrikosov} and other works on vortices in rotating Helium \cite{yarmchuk, donnelly}, it has been widely verified that rotating Bose-Einstein condensates (BEC), under appropriate angular velocities of rotation\cite{abo, madison, cooper, schweikhard}, indeed show a triangular vortex lattice as the most stable state. Alternative ways have been devised to create vortices, such as phase imprinting\cite{leanhardt, machida, machida2, azure} and, very recently, by creation of artificial gauge fields\cite{spielman, dalibard} and by varying in time the fields of a magnetic trap\cite{vanderlei}. In this article we introduce another way of creating vortices; one that necessarily requires the presence of more than one atomic internal spin state coupled to an external transverse magnetic field.  In this regard it is of relevance to recall that, even though very early in the development of optical traps\cite{Stenger,Miesner} it was pointed out\cite{Ho,Ohmi,Yip} the additional richness of spinor or multicomponent Bose-Einstein condensates, most of the efforts in creating and studying vortices were focused on rotating condensates\cite{Fetter}. The novelty of our proposal, in addition to exploit the spinorial structure of the condensates, is that the vortices can be created essentially at any desired spatial location. This property, as we shall see, depends on the ability to produce a corresponding on-demand tailored magnetic field.  Although we have tested our ideas with 2 and 3D BEC clouds, here for simplicity, we shall limit ourselves to analyze a 2D case only.

The essence of vortex appearance follows from essentially the same argument provided by Abrikosov\cite{abrikosov}. In superconductors, vortices appear in the spatial points where the type II superconductor allows the magnetic field to penetrate. In those vortices (or flux tubes as they are also known) the density is zero and the superconductor current circulates around the vortex to cancel the field in the bulk. In our proposal the situation is very similar, except that we work with multi-component wavefunctions and, unlike Abrikosov, we work with non-uniform magnetic fields. 

We find, interestingly, that the vortices are nucleated not where the magnetic field is able to penetrate the condensate, but on the spatial points where the field is zero. Additionally, the vortex solution will not have the same charge in all spin components of the wave function since, as we show, the difference in charges of two adjacent spin components must obey a selection rule dictated by the way in which the magnetic field vanishes. Moreover, we find that always one spin component must have no vortex, i.e. $Q=0$, and thus, such a  spin component will develop a density accumulation or spike so as to cancel the vortex formed on the other component(s), leaving the total density unaltered in its Thomas-Fermi-like form. In this way, vortices will be nucleated ``on demand" at the spatial positions where the magnetic field is zero, albeit not with the same charge on all spin components. 

This approach presents some additional interesting consequences. The first one is that vortices can be created with arbitrary integer charge, $Q = 1,2,3 \dots$, by selecting a field with appropriate multipolar order and by including or not an axial magnetic field, that is, perpendicular to the plane of the 2D BEC cloud. In this way, not only the position of the vortices, but their charge as well, can be controlled using on-demand tailored magnetic fields. It is of relevance to point out here that if the axial field is zero, the spin components that show vortices are chosen by an spontaneous symmetry breaking mechanism, while the presence of such a field breaks the symmetry and one can predict both the charge and the component where the vortices will be nucleated.

The second interesting consequence of our approach is that it is possible to assemble vortex lattices ``on demand" by selecting magnetic fields with zeroes on the lattice sites. Now, since all solutions obtained from the Gross-Pitaevskii and Ginzburg-Landau equations are of minimum free energy, we conclude that triangular vortex lattices observed both at superconductors \cite{essmann} and BEC alike \cite{abo, madison, cooper, schweikhard}, are a peculiarity consequence of the use of \textit{uniform} magnetic fields. Rotating BEC are not an exception to this. It is possible to see that a rotating BEC is equivalent to a superfluid in a \textit{uniform} magnetic field. In this way, our approach may have the further advantage of providing evidence of a unique cause underlying the different vortex nucleation techniques currently at use.

In this article we first analyze the corresponding Gross-Pitaevskii (GP) equation to describe how the vortices appear. The discussion is valid in principle for any spin value of the atoms composing the BEC state, namely, $j =1/2, 1,3/2, \dots$. We discuss a very general solution for any transverse magnetic field that vanishes linearly, quadratically, etc. at chosen locations. We shall see that the TF solution guides the formation of vortices, including its size and location if the axial field is nonzero. In Section III, we present a variety of different cases by numerically solving the corresponding GP equations, in order to illustrate our results.

 In the last section we suggest how this situation may be experimentally realized in a gas of $^{87}$Rb occupying all three $F = 1$ states in an optical trap. In this regard, we point out that precursors of the present study are the analysis of Refs. \cite{Bulgakov} and \cite{Kasamatsu}, the former discussing the appearance of vortices in the center of a Ioffe-Pritchard trap in a $F =1$ condensate, while the latter being a detailed study of vortices on two-level condensates, yet, in rotating systems.

\section{A multicomponent BEC in a {\it transverse} magnetic field}

Consider a 2 or 3 dimensional BEC superfluid at zero temperature.  The gas is confined by an optical trap. In addition, there exists an external {\it transverse} magnetic field that couples to the $x$ and $y$ spin components. The atoms have spin $j$. The $2j +1$ GP equations in a very general form are,
\begin{eqnarray}
&&\left[ -\frac{\hbar^2}{2M} \nabla^2 + V_{ext}^m(\vec r) + \sum_{m^\prime} 
 g^{m m^\prime} \left| \psi_{m^\prime}(\vec r)\right|^2 \right] \psi_m(\vec r) \nonumber \\
 && - m_0\sum_{m^\prime} \left[J_x^{ m m^\prime} B_x(\vec r)+J_y^{m m^\prime}B_y(\vec r) +  J_z^{m m^\prime} B_z\right] \psi_{m^\prime}(\vec r) \nonumber\\
 &&= \mu \psi_{m}(\vec r),
 \label{GP-gral}
\end{eqnarray}
where $\vec r$ is the $D$-dimensional position vector. The indices, $m, m^\prime = -j, -j + 1, \dots, j-1 ,j$, denote the $2j + 1$ different spin components. The confining external potentials are $V_{ext}^m(\vec r)$ and we have considered the possibility that each spin component feels a different confining trapping potential; however, for illustration purposes we shall assume below that they are all the same harmonic potential. In general, we assume the external trapping potentials to be smooth at any point. We can also, in principle, assume that the scattering length is different for any pair of components; but again, in the numerical evaluations below, we will consider the simple case $g^{\alpha \beta} = g \delta_{\alpha \beta}$. This is certainly an important point since it is known\cite{Klausen} that the scattering length truly depends on the internal atomic states and this in turn leads to different types of macroscopic ``magnetic" properties of the condensates. By assuming a unique positive scattering length throughout, one may assert that the condensates with which we are dealing are at the border of being ferromagnetic or polar\cite{Ho}. This does not invalidate the point we are highlighting here.

The most important part, and the novelty of our argument, is the coupling of the spin components via an  external {\it inhomogeneous} transverse magnetic field $\vec B(\vec r) = B_x(\vec r) \hat{e}_x + B_y(\vec r) \hat{e}_y$. Accordingly, $J_x^{m m^\prime}$ and $J_y^{m m^\prime}$ are the $x$ and $y$ angular momentum matrices of spin $j$. In addition, we also include a {\it uniform} axial magnetic field $B_z$. In (\ref{GP-gral}),  $\mu$ is the chemical potential and $m_0$ the magnetic moment of an atom.

Now, the external magnetic field must obey $\nabla \cdot \vec B = 0$ at any point. In addition, since we want to resemble a true experimental situation, we should demand that throughout the condensate $\nabla \times \vec B = 0$ is also obeyed.  In what follows we will consider these two restrictions to build the proposed fields, however, we will also include more general situations. We also defer to the following sections the question of how the proposed fields may be produced. 

The main point of the discussion is based on proposing that the transverse magnetic field becomes zero at isolated points $\vec r_0$, namely, $B_x(\vec r_0) = B_y(\vec r_0) = 0$. Placing the origin at $\vec r_0$, the magnetic field very near that point obeying $\nabla \cdot \vec B = 0$ and $\nabla \times \vec B = 0$, have the following form,
\begin{equation}
B_x (x,y) \approx {\cal B}_l  r^l  \cos l \phi \label{B1}
\end{equation}
and
\begin{equation}
B_y (x,y) \approx - {\cal B}_l r^l  \sin l \phi \label{B2}
\end{equation}
where $l = 1,2,3 \dots$, $\tan \phi = y/x$ and $r^2 = x^2 + y^2$. We shall call $l =1$ dipolar, $l = 2$ quadrupolar, etc. following the usual electrodynamics notation\cite{electroGreiner}. ${\cal B}_l$ are constants with appropriate units. The $z$ component of the field $B_z$ is assumed constant. For the sake of exemplifying the two simplest cases, we write down the explicit forms for $l = 1$ and $l = 2$,
\begin{equation}
\vec B (x,y) \approx {\cal B}_1 \left( x \hat{e}_x - y \hat{e}_y \right) \>\>\>\>{\rm for } \>\>\>\> l = 1\label{B11}
\end{equation}
and
\begin{equation}
\vec B (x,y) \approx  {\cal B}_2 \left( (x^2 - y^2) \hat{e}_x - 2xy \hat{e}_y\right) \>\>\>\>{\rm for } \>\>\>\> l = 2  \label{B12}.
\end{equation}

Let us turn now to the main problem: vortex nucleation using the fields already discussed. Without loss of generality, we assume the same confining potential $V_{ext}(\vec r)$ for all components and the same strength $g$ for all interacting pairs. We consider the external magnetic  fields $B_x$ and $B_y$ actually given by (\ref{B1}) and (\ref{B2}), but centered at the origin $\vec r_0 = 0$. Below we shall discuss the conditions for the location of of the zeroes of the fields to be set anywhere within the condensate. Using the standard properties of angular momentum\cite{Cohen}, equations (\ref{GP-gral}) reduce to the coupled set,
\begin{eqnarray}
\label{GP-m}
&&\left[ -\frac{\hbar^2}{2M} \nabla^2 + V_{ext}(\vec r) + g \sum_{m^\prime} 
 \left| \psi_{m^\prime}(\vec r)\right|^2 - \mu + m \> m_0 B_z \right] \psi_m(\vec r)\nonumber   \\
 && + \frac{1}{2}  m_0 {\cal B}_l r^l \left[\sqrt{(j+m)(j-m+1)} e^{il\phi} \> \psi_{m-1}(\vec r)\right.
 \nonumber\\
 &&\left.+\sqrt{(j-m)(j+m+1)} e^{-il\phi} \> \psi_{m+1}(\vec r) \right]  = 0
\end{eqnarray}
for all $m = -j, -j + 1, \dots, j-1, j$ and with the convention $\psi_{j+1} = \psi_{-j-1} \equiv 0$. One proposes a solution of the type,
\begin{equation}
\psi_m (\vec r) = f_m(r) \> e^{i\zeta_m \phi} \label{sol}
\end{equation}
where $\zeta_m$ take integer values in order to have single-valued solutions. Substitution of (\ref{sol}) into (\ref{GP-m}) yield the condition,
\begin{equation}
\zeta_m - \zeta_{m-1} = l ,\label{confas}
\end{equation}
for $m = j, j-1, \dots, -j +1$. The remaining
$f_m(r)$ are real functions  obeying,
\begin{eqnarray}
\label{GP-f}
&&\left[ -\frac{\hbar^2}{2M} \left(\frac{1}{r}\frac{\partial}{\partial r}r \frac{\partial}{\partial r}- \frac{\zeta_m^2}{r^2}\right)  + V_{ext}(r) 
\right.\nonumber\\
&&\left.+ g \sum_{m^\prime} 
  f_{m^\prime}^2(r) - \mu + m \> m_0 B_z \right] f_m( r)  \\
 && + \frac{1}{2}  m_0 {\cal B}_l r^l \left[\sqrt{(j+m)(j-m+1)}  f_{m-1}( r) \right.\nonumber
 \\
&& \left.+\sqrt{(j-m)(j+m+1)} f_{m+1}(r) \right]  = 0 .\nonumber
\end{eqnarray}

An important aspect to be taken into account here, and numerically verified below, is the fact that since is one is dealing with a macroscopic gas, the atomic interaction term is ``large" (i.e. $g$ is proportional to the number of atoms) compared the kinetic energy term (proportional to the Laplacian). Neglecting the kinetic energy term yields the so-called Thomas-Fermi solution, but, it is the kinetic term the responsible for the emergence of vortex solutions. Further, as we now discuss, there appear to be two different types of vortices depending on whether the axial field component $B_z$ is present or not.

Let us consider the Thomas-Fermi (TF) solution obtained by neglecting the kinetic energy term in (\ref{GP-f}). One first finds that the {\it total} particle density
\begin{equation}
\rho(r) = \sum_{m^\prime} f_{m^\prime}^2(r)
\end{equation}
obeys the following $2j+1$ different solutions,
\begin{equation}
\rho(r) = \frac{1}{g} \left(\mu - V_{ext}(r) + m m_0 \sqrt{B_z^2 + {\cal B}_l^2 r^{2l}}\right)
\end{equation}
for $m = -j, -j + 1, \dots, j-1, j$. Given all variables equal, it appears that $m = j$, its largest value, yields the lowest value of the chemical potential $\mu$. Since this is the eigenenergy of the GP equation, we take the TF solution as given by,
\begin{equation}
\rho_{TF}(r) = \frac{1}{g} \left(\mu - V_{ext}(r) + j m_0 \sqrt{B_z^2 + {\cal B}_l^2 r^{2l}}\right) .\label{TF}
\end{equation}
As we show below, we always find that the total density agrees essentially with this solution, except at the edges of the cloud. Notice that it does not show any trace of the presence of a vortex. Substitution of the solution (\ref{TF}) into (\ref{GP-f}) yields a set of $2j+1$ coupled linear equations for $f_m(r)$. These equations can be explicitly solved for $j = 1/2$, $j = 1$, $j = 3/2$, etc. One finds two very different behaviors depending on the values of $B_z$.

First, we analyze $B_z = 0$. One can check case by case the solutions.
For $j = 1/2$,
\begin{equation}
f_{1/2}^2(r) = f_{-1/2}^2(r) = \frac{1}{2} \rho_{TF}(r) .
\end{equation}
For $j = 1$,
\begin{equation}
f_{1}^2(r) = f_{-1}^2(r) = \frac{1}{4} \rho_{TF}(r) \>\>\>\>\>\>\>\> f_{0}^2(r) = \frac{1}{2} \rho_{TF}(r) .
\end{equation}
For $j = 3/2$,
\begin{eqnarray}
&&f_{3/2}^2(r) = f_{-3/2}^2(r) = \frac{1}{8} \rho_{TF}(r)\nonumber \\
&&f_{1/2}^2(r) = f_{-1/2}^2(r) = \frac{3}{8} \rho_{TF}(r) ,
\end{eqnarray}
and so on. That is, the particle density of any component is a fraction of the total density, and the density distribution is symmetric. Since all density components are proportional to the TF solution, none of them shows a vortex. However, the Laplacian term induces a vortex solution at the origin. That is, near the origin where the magnetic field vanishes, the density components behave as,
\begin{equation}
\psi_m(r,\phi) \approx A_m r^{\zeta_m}\> e^{i\zeta_m \phi} \label{vortex},
\end{equation}
where $A_m$ are amplitude and normalization constants and $\zeta_m$ must be an integer, including zero, by demanding single valuedness of the wave functions $\psi_m(r,\phi)$. These satisfy,
\begin{equation}
\left(\frac{1}{r}\frac{\partial}{\partial r}r \frac{\partial}{\partial r}- \frac{\zeta_m^2}{r^2}\right) \psi_m(r,\phi) \approx 0
\end{equation}
{\it independently} of the value of $\zeta_m$. If $\zeta_m \ne 0$, the solution (\ref{vortex}) is a vortex of topological charge, or order,  $Q = \zeta_m$ and of size $\xi = \sqrt{\hbar^2/2M\mu}$, namely, the size of usual Gross-Pitaevskii vortices\cite{abrikosov,gross,pitaevskii}. We recall that the superfluid velocity is given by 
\begin{equation}
\vec v_s = \frac{\hbar}{M} \zeta_m \nabla \phi
\end{equation}
such that the circulation around the vortex is given by,
\begin{equation}
{\cal C}_m = \oint \vec v_s \cdot d\vec r = 2 \pi \frac{\hbar}{M}  \zeta_m .\label{circ}
\end{equation}
In summary, near the zero of the magnetic field, $r \le \xi$, the kinetic energy term does not contribute because of the appearance of a vortex, while for $r > \xi$ the TF solution takes over. 

An interesting problem now arises. Because of the symmetry of the Hamiltonian, one cannot assert in which component or components, the vortices may appear. An spontaneous symmetry-breaking mechanism must enter to decide it. There are two restrictions, one is the rule of the phases given by (\ref{confas}) and the other the fact that the {\it total} density follows TF everywhere. The latter imply that at least one component cannot show a vortex. The former thus gives us the possible solutions. For $j = 1/2$,
\begin{equation}
Q_{1/2} = l \>\>\>Q_{-1/2} = 0 \>\>\>{\rm or}\>\>\> Q_{1/2} = 0 \>\>\>Q_{-1/2} = -l  .
\end{equation}
As mentioned, both solutions are possible, the system takes one by an spontaneous symmetry-breaking mechanism. This result has been numerically verified. For $j=1$,
\begin{eqnarray}
&&\phantom{{\rm or}}\>\>Q_{1} = 1 \>\>\>  Q_{-1} = -1 \>\>\>Q_{0} = 0 
\nonumber\\
&&{\rm or}\>\> Q_{1} = 2 \>\>\>Q_{0} = 1\>\>\>\>\>\>\>\>\>Q_{-1} = 0 
\nonumber\\
&&{\rm or}\>\>
Q_{1} = 0 \>\>\>Q_{0} = -1 \>\>\> \>\>\>Q_{-1} = -2 .\label{j1ssb} 
\end{eqnarray}
For $j = 3/2$,
\begin{eqnarray}
&&\phantom{{\rm or}}\>\> Q_{3/2} = 2 \>\>Q_{1/2} = 1 \>\>Q_{-1/2} = 0 \>\>\>\>\> Q_{-3/2} = -1 \nonumber \\
 &&{\rm or}\>\> Q_{3/2} = 1 \>\>Q_{1/2} = 0 \>\>Q_{-1/2} = -1 \>\> Q_{-3/2} = -2.\nonumber
 \\
 &&
 \end{eqnarray}
 
 The presence of an axial constant field $B_z \ne 0$ breaks the symmetry and yields only one of the above solutions. However, it does affect the sizes of the vortices as well. That is, if $B_z \ne 0$, the TF solutions already show a vortex solutions. Let us return to the set of equations (\ref{GP-f}) with $\rho_{TF}(r)$ given by (\ref{TF}). Take $B_z > 0$ for definiteness, then it is easy to verify that in the vicinity of the zero of the transverse magnetic field, the TF density components behave as, for $j = 1/2$
 \begin{eqnarray}
&& f_{1/2}^2(r) \approx  \rho_{TF}(0) \left(\frac{{\cal B}_l}{2 B_z}\right)^2 r^{2l}\nonumber\\
&&  f_{-1/2}^2(r) \approx \rho_{TF}(0) \left(1 -\left(\frac{{\cal B}_l}{2 B_z}\right)^2 r^{2l} \right) \label{j12bz}
\end{eqnarray}
and, in turn, this implies $\zeta_{1/2} = l$ and $\zeta_{-1/2} = 0$. That is, while there is a vortex of charge $l$ at the $+1/2$ component, there is a ``density spike" at the $-1/2$ component. If $B_z <0$, then the spike is at the $+1/2$ component, while a vortex of charge $-l$ appears at the $-1/2$ one. The size of the vortex is now $\xi = (B_z/{\cal B}_l)^{1/l}$. All this is numerically verified below. 

For completeness, we also show the vortices solution for $j = 1$, for $B_z > 0$. These are, 
\begin{eqnarray}
&&f_{1}^2(r) \approx  \rho_{TF}(0)\left(\frac{{\cal B}_l}{2 B_z}\right)^4 r^{4l} \nonumber\\ 
&&f_{0}^2(r) \approx 2 \rho_{TF}(0)\left(\frac{{\cal B}_l}{2 B_z}\right)^2 r^{2l} \nonumber \\
&&f_{-1}^2(r) \approx  \rho_{TF}(0)\left[ 1 - 2\left(\frac{{\cal B}_l}{2 B_z}\right)^2 r^{2l}- \left(\frac{{\cal B}_l}{2 B_z}\right)^4 r^{4l} \right] \nonumber\\
&&\label{j11bz}
\end{eqnarray}
which imply that $\zeta_1 = 2 l$, $\zeta_0 = l$ and $\zeta_{-1} = 0$, that is, there is vortex of charge $+2l$ at the $+1$ component, a vortex of charge $+l$ at the $0$ component, and a density spike at the $-1$ component. Again, the size of the vortices is $\xi = (B_z/{\cal B}_l)^{1/l}$. If the axial field is negative $B_z <0$, the solution is a spike in the $+1$ component, a vortex $-1$ and a vortex $-2$ at the $0$ and $-1$ components.

Now, let us note that since the above argument only gives the solution at the given point $\vec r_0$ where the external magnetic field vanishes, one can consider cases where a magnetic field has zeroes at different spatial locations. Applying the same argument near any of those points, a vortex will be nucleated at any desired position. In the following section we exemplify all these features.

\section{A gallery of vortices on demand}

In this section we illustrate the present mechanism for vortex formation with a variety of cases. For simplicity, several assumptions are made. The trapping potential is assumed to be an isotropic harmonic potential for all spin components with the same frequency $\omega$. The common scattering length $g$ is  a free parameter.  
Units in which $\hbar = m = \omega = 1$ are used. In all cases, the corresponding set of GP-equations is numerically solved with the method proposed by Zeng and Zhang\cite{zeng}. The chemical potential is part of the solution and it is also our convergence parameter. Its accuracy is at least 1 part in $10^5$, namely, beyond the significant figures we show in all cases. Additional details are provided case by case.

In most of the figures below we show, (a) the particle densities of the different components $\rho_m(x,y) = |\psi_m(x,y)|^2$, for $m = -j, -j+1, \dots, j-1, j$,  in the form of color coded density plots; (b) velocity fields  $\vec v_m = \frac{\hbar}{M} \nabla \Phi_m$ where $\tan \Phi_m = {\rm Im} \psi_m/{\rm Re} \psi_m$; and (c) the circulation, given by 
\begin{equation}
{\cal C}_m = \oint \vec v_m \cdot d\vec l .
\end{equation}
For numerical simplicity, we use integration circuits in the form of squares of length side $r$ centered at the zeroes of the transverse magnetic fields. We increase the size of the square until the circulation converges.
\begin{figure}
\begin{center}
\includegraphics[width=0.47\textwidth]{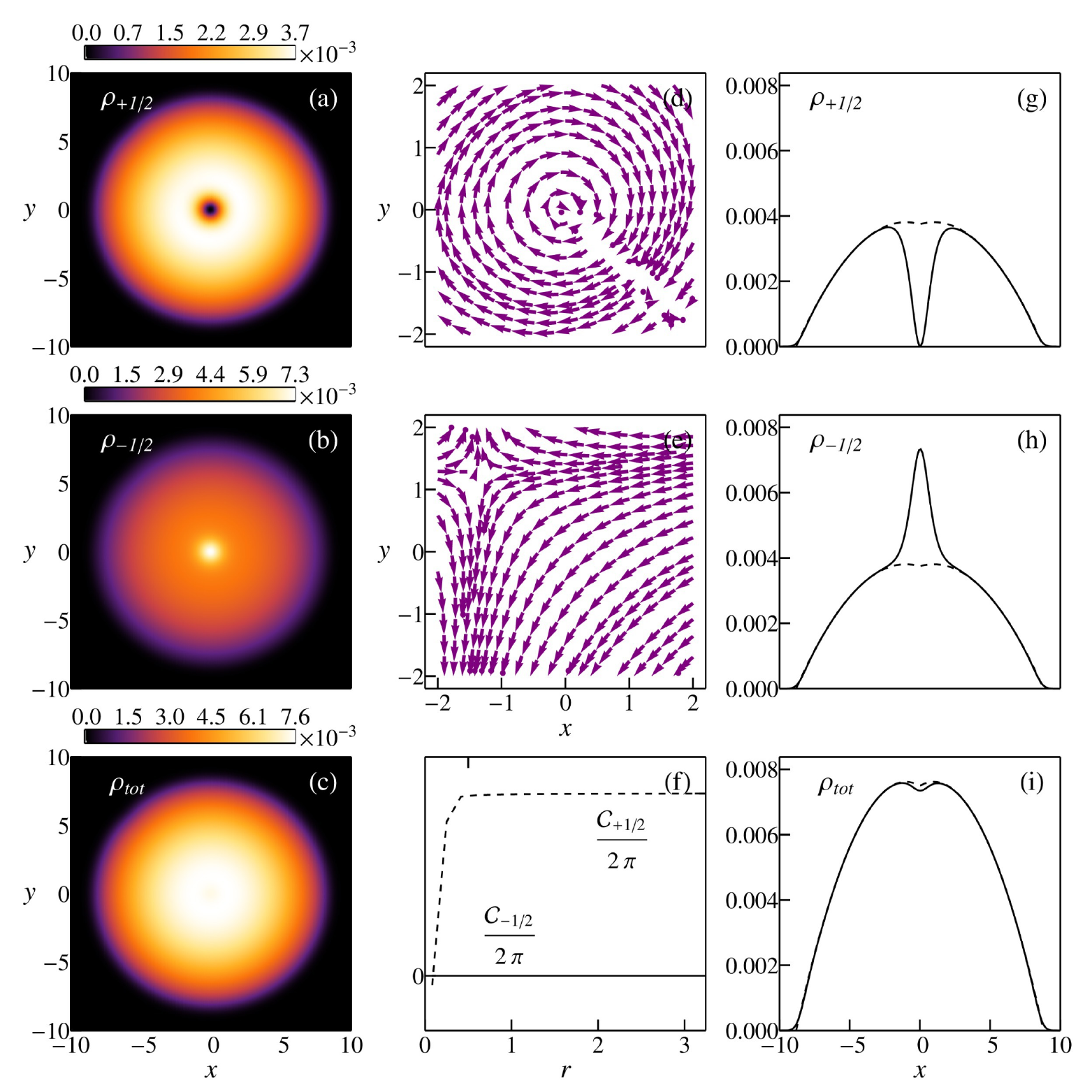}
\end{center}
\caption{(Color online) Vortex at the origin in a  $j = 1/2$ BEC produced by a dipolar $l = 1$ transverse magnetic field with $B_z = 0$. The $m = 1/2$ component shows a vortex of charge $Q = 1$, while the $m = -1/2$ component presents a density spike. We show density plots, $\rho_{\pm 1/2}$ in (a) and (b), and (c) the total density. Velocity fields $\vec v_{\pm 1/2}$ in (d) and (e), and (f) is their circulation around the zero of the field. (g) to (i) are comparison of GP calculations, solid lines, with TF analytic solutions, dotted lines. Parameters are $g = 4000$, $m_0 {\cal B}_1 = 1.0$. Chemical potential is $\mu=29.96$.}
\label{F1}
\end{figure}

\begin{figure}
\begin{center}
\includegraphics[width=0.47\textwidth]{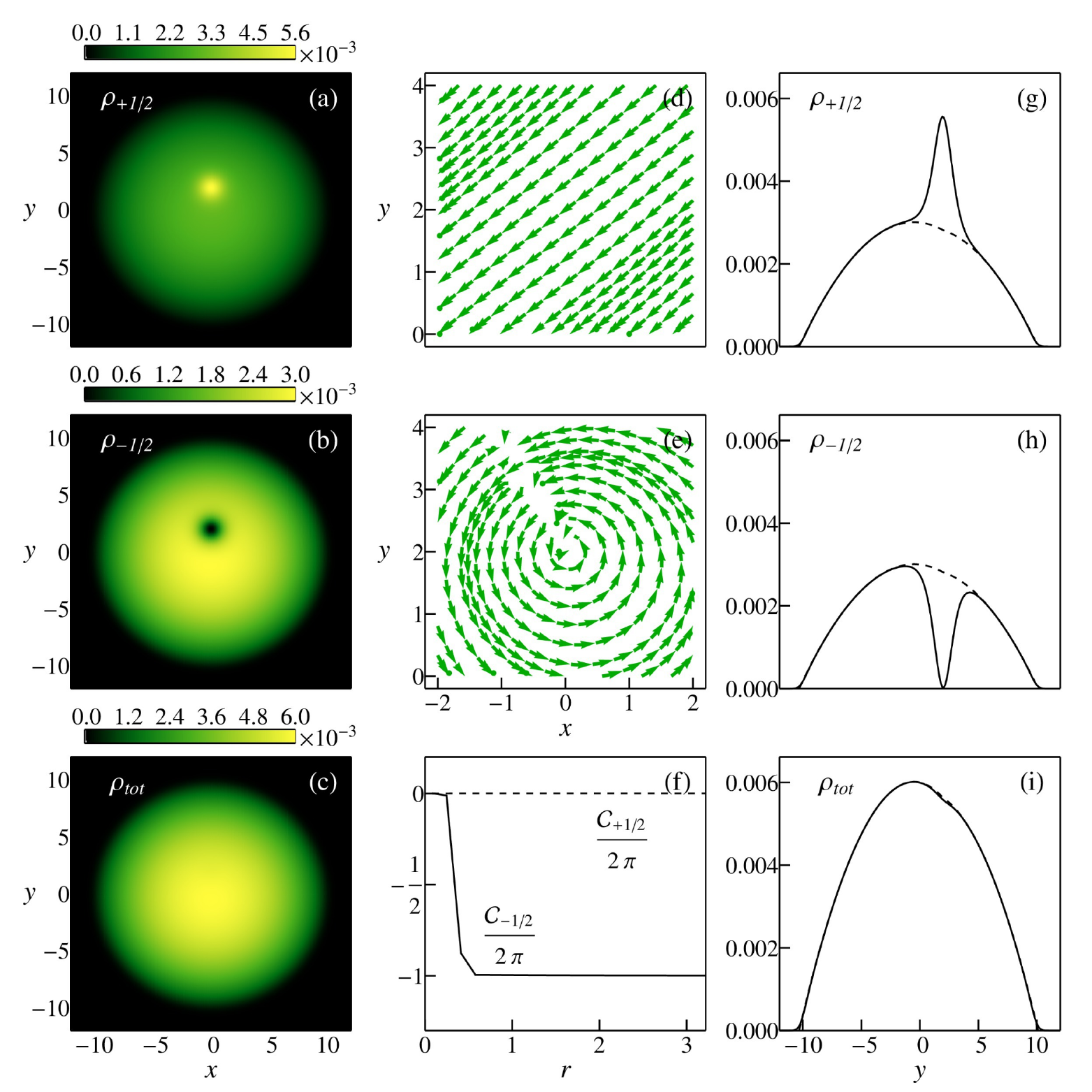}
\end{center}
\caption{(Color online) Vortex in a $j = 1/2$ BEC, created at the chosen vanishing place $\vec r_0 = (0,2)$ of a transverse dipolar $l = 1$ magnetic field with $B_z = 0$. A vortex of charge $Q = -1$ appears in the $m = -1/2$ component. The $m = 1/2$ component shows a density spike. The panels (a) to (f) are described in figure caption of Fig. \ref{F1}. Parameters are: $g=8000$,  $m_0 {\cal B}_1 = 0.5$. Chemical potencial is $\mu=47.00$.}
\label{F2}
\end{figure}

\subsection{Single vortices, $B_z = 0$.}

This is the simplest case to consider and we begin with two internal states, namely $j = 1/2$, zero axial field $B_z = 0$ and a transverse magnetic field $(B_x,B_y)$ that vanishes at the origin in a dipolar fashion. Letting $l =1$ in Eqs.(\ref{B1}) and (\ref{B2}), the field is dipolar in the sense that it is a linear combination of spherical harmonics of order one, see Eq. (\ref{B11}). It is very similar to the Ioffe-Pritchard trap field except that our field has a constant $z$ component and, consequently, the sign of the $y$ component must change to preserve Maxwell's equations.

Fig. \ref{F1} shows a vortex at the origin $\vec r_0 =0$ and Fig. \ref{F2} a vortex at an arbitrary $\vec r_0$. Panels (a)-(b) show density plots of components $\rho_{1/2}$ and $\rho_{-1/2}$, while (c) shows the total density $\rho = \rho_{1/2} + \rho_{-1/2}$; panels (d)-(e) show the velocity fields $\vec v_{1/2}$ and $\vec v_{-1/2}$, while (f) the corresponding circulations ${\cal C}_{1/2}$ and ${\cal C}_{-1/2}$; figures (g)-(h)-(i) show the comparison of the predicted TF solution, with dotted lines, with the actual calculations for both the components and the total densities, solid lines. The particular numerical values of the parameters are given in the figure captions.

Fig. \ref{F1} shows a vortex of charge $Q=+1$ in the $m=+1/2$ spin component while a ``spike" in the $m=-1/2$ one at the origin. The total density does not register the presence of the vortex. That is, the vortex and the spike cancel each other as the corresponding density components are summed up. The appearance of the vortex is also clearly seen in the velocity fields and the value of the charge follows from the circulation. Moreover, the density field of the vortex at $m = 1/2$ does show a quadratic profile as expected, with its width scaling with the Abrikosov size.

It is very interesting to note that the TF solutions, dotted lines in panels (g)-(h)-(i) of Fig. \ref{F1}, do not show a vortex and, thus, the predicted TF integrated densities,
\begin{equation}
N_{\pm 1/2} = \int \rho_{\pm 1/2}(x,y) dx dy
\end{equation}
yield $N_{\pm 1/2}/N = 1/2$ exactly, while the full numerical GP solutions deviate slightly from 1/2 because of the appearance of the vortex. That is, the population $N_{1/2} = 0.49 N$ of the vortex component has a slightly smaller value than 1/2, while such a deficit appears as an excess  in the spike component $N_{-1/2} = 0.51 N$. The total density indeed follows quite closely the TF solution. All these observations are in agreement with the theoretical predictions of the previous Section. 
\begin{figure}
\begin{center}
\includegraphics[width=0.47\textwidth]{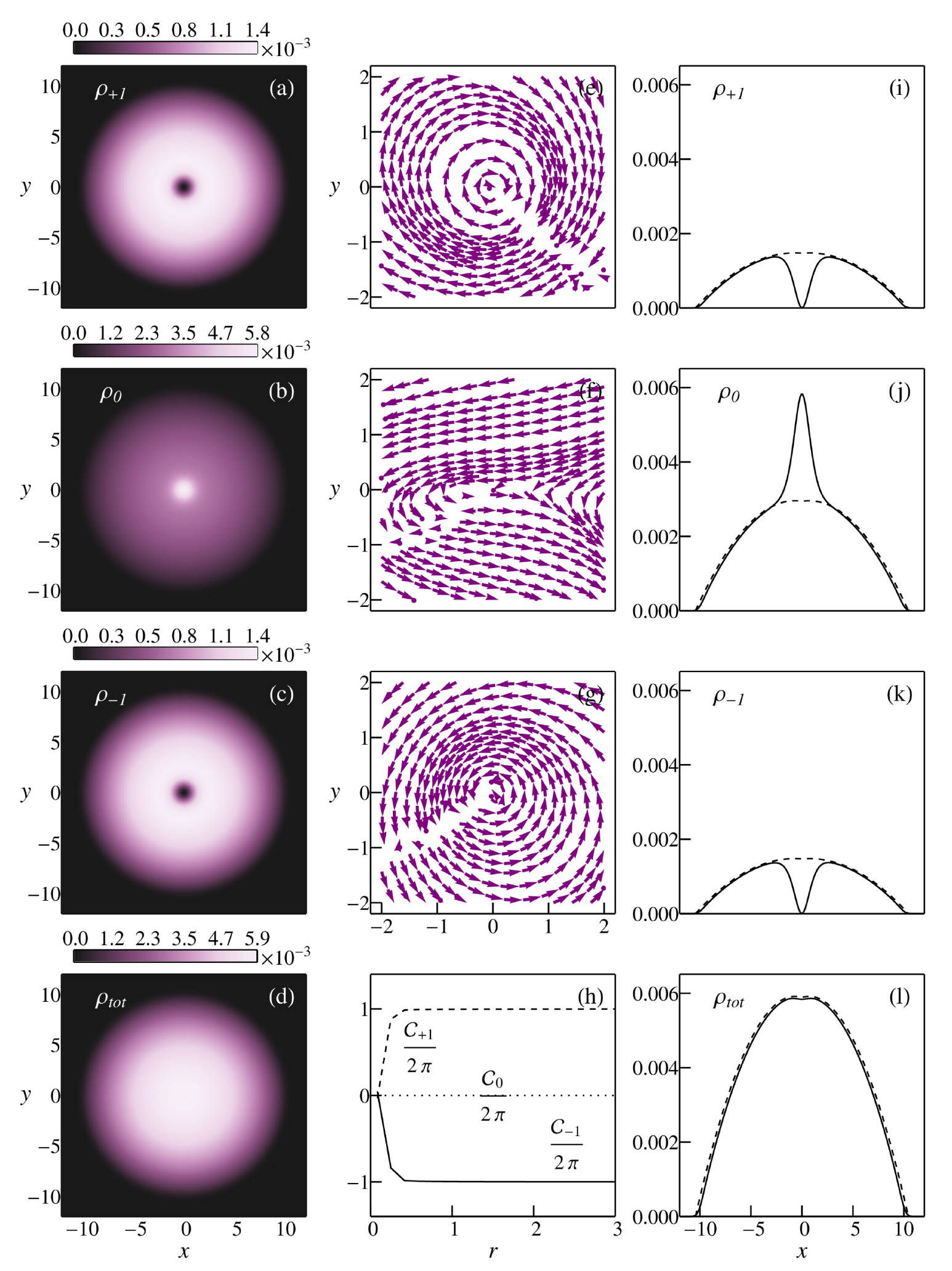}
\end{center}
\caption{(Color online) Vortices at the origin in a $j = 1$ BEC, generated by a dipolar $l=1$ transverse magnetic field with $B_z = 0$. There appear vortices of charge $Q = \pm 1$ at the $m = \pm 1$ components respectively. The component $m = 0$ shows a density spike. Panels (a) to (c) are density plots $\rho_m$, while (d) is the total one. (e) to (g) are velocity fields $\vec v_m$, while (h) is their circulations. (i) to (l) are comparison of GP calculations, solid lines, with TF analytic solutions, dotted lines. Parameters are $g = 8000$, $m_0 {\cal B}_1 = 0.5$. Chemical potential is $\mu=47.10$}
\label{F3}
\end{figure}

\begin{figure}
\begin{center}
\includegraphics[width=0.47\textwidth]{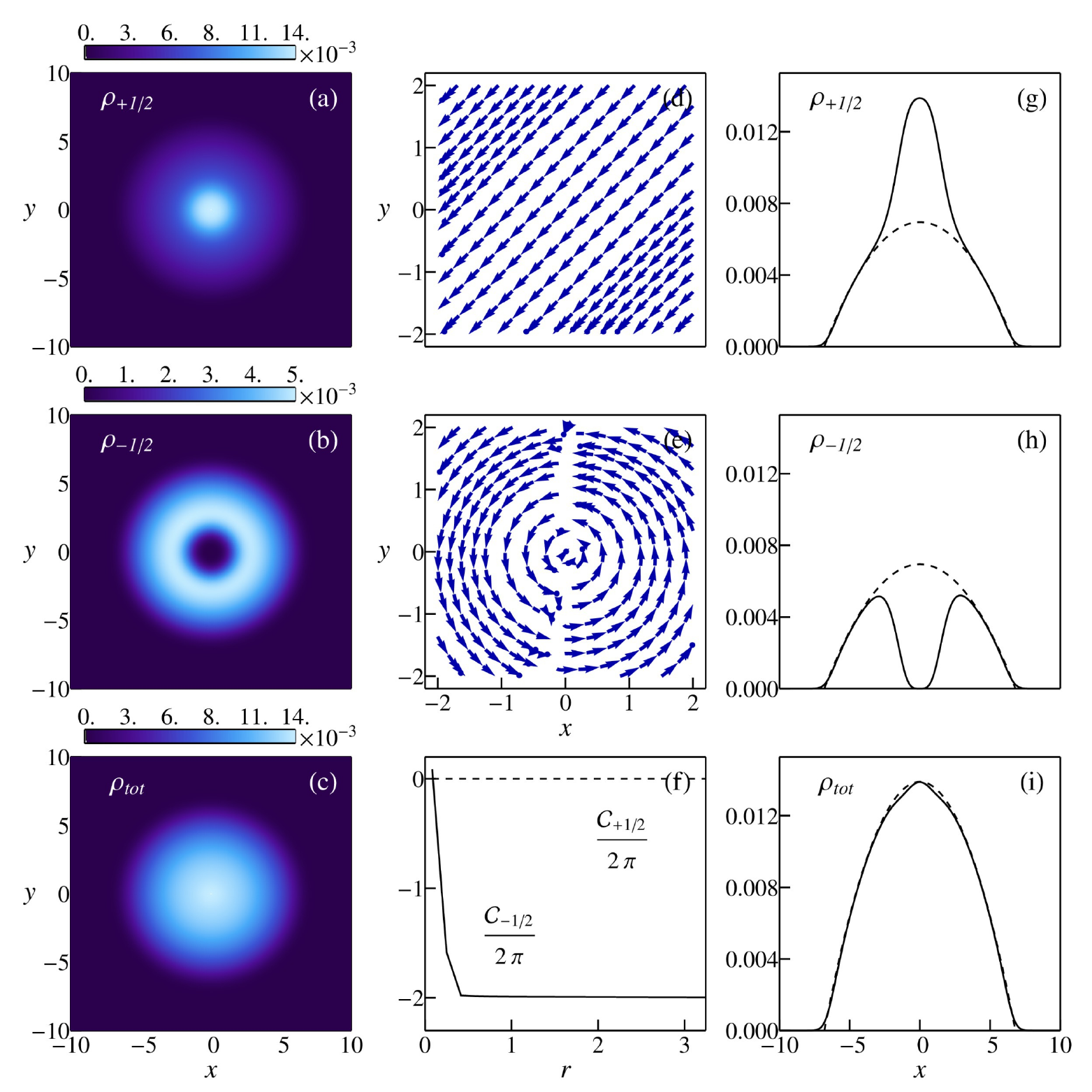}
\end{center}
\caption{(Color online) Vortex at the origin in a $j = 1/2$ BEC produced by a quadrupolar $l=2$ transverse magnetic field with $B_z = 0$. The vortex is in the $m = -1/2$ component has charge $Q = -2$. The $m = 1/2$ component shows a density spike. The panels (a) to (f) are described in figure caption of Fig. \ref{F1}. Parameters are: $g=1000$,  $m_0 {\cal B}_1 = 0.2$. Chemical potencial is $\mu=13.89\pm0.3$.}
\label{F4}
\end{figure}

We also observe the spontaneous symmetry breaking  mechanism in this case due to the zero $B_z$ field. In the case shown in Fig. \ref{F1} the vortex appears in the $m= +1/2$ component. We have repeated the calculation several times, with the same value of the parameters, except changing the initial condition with random values of the densities components, and we have found that in nearly half of the cases the stable vortex is just as in Fig. \ref{F1} while in the other half the situation is reversed: a vortex of charge $Q = -1$ appears now in the $m=-1/2$ component with all the density values simply inverted with respect to those of Fig. \ref{F1}.

Figure \ref{F2} shows the same situation as in Fig. \ref{F1} except that the vortex was created at the position $\vec r_0 = (0,2)$, which is where the transverse magnetic field was made to vanish $B_x(\vec r_0) = B_y(\vec r_0) =0$. This illustrates one of the possibility of creating vortices at any desired location. It is of interest to point out that the chemical potential in this case is quite greater than in the case of Fig. \ref{F1}. This is because the atom density  is globally given by the Thomas-Fermi contribution, see Eq. (\ref{TF}), and this depends strongly on the interaction parameter $g$. Note that $g = 4000$ in the case of Fig. \ref{F1} while it is $g = 8000$ in Fig. \ref{F2}. 
 
Figure \ref{F3} shows the appearance of vortices of charge $Q = \pm 1$, in a multicomponent BEC with $j = 1$ with no axial field $B_z = 0$ but with a transverse  field $(B_x,B_y)$ that becomes zero at the origin with a dipolar $l =1$ behavior. That is, there are now three internal components $m = 1, 0, -1$. Similarly to the previous figures, the panels (a)-(b)-(c) show density plots $\rho_m$, with (d) the total one $\rho = \rho_1+\rho_0+\rho_{-1}$; panels (e)-(f)-(g) show velocity fields $\vec v_m$ and (h) their circulations ${\cal C}_m$. Panels (i)-(j)-(k)-(l) show comparisons of actual calculations, solid lines, with TF solutions, dotted lines.

As predicted in Section II, the lack of $B_z$ field yields the possibility of vortices in three cases different cases for $j = 1$, see Eq. (\ref{j1ssb}), namely vortices $Q = \pm 1$ at $m = \pm 1$ and no vortex at $m =0$; vortex $Q = 2$ at $m = 1$, $Q = 1$ at $m = 0$ and no vortex at $m = -1$; and no vortex at $m = 1$ while $Q = -1$ and $Q = -2$ at $m = 0$ and $m = -1$ respectively. We have found all three cases with a rate of a third of the times for each case. Figure \ref{F3} shows the first of these possibilities, with all the calculated quantities in agreement with the theory of Section II. According to TF solution, the population should be $N_1 = N_{-1} = N/4$ and $N_0 = 1/2$. The actual solution is close to these values, $N_1 = N_{-1} = 0.23 N$, $N_0  =.54 N$, again with an equal deficit in the components with vortices and an excess in the component with the spike. The sum does agree with the full TF solution. We point out that some features of this particular case have also been discussed in Ref. \cite{Bulgakov}

We now turn our attention to the vortex creation with no axial field, $B_z = 0$, and a transverse field $(B_x,B_y)$ that vanishes at the origin in a {\it quadrupolar} form, namely $l = 2$, see Eq.(\ref{B12}). This is illustrated in Fig. \ref{F4}, where a BEC of $j = 1/2$ is studied. The panels are arranged just like in Fig. \ref{F1}, but now a vortex of charge $Q = -2$ is found in the $m = -1/2$ component and no vortex in the $m = 1/2$ one. The fact that it is charge $Q = - 2$ follows both from the circulation calculation, panel (f), and from the size and form of the vortex as shown in panel (g): the vortex scales as $r^{2Q}$ at its location. This is more clearly by comparing with the figure \ref{F1}(g). 

\begin{figure}
\begin{center}
\includegraphics[width=0.47\textwidth]{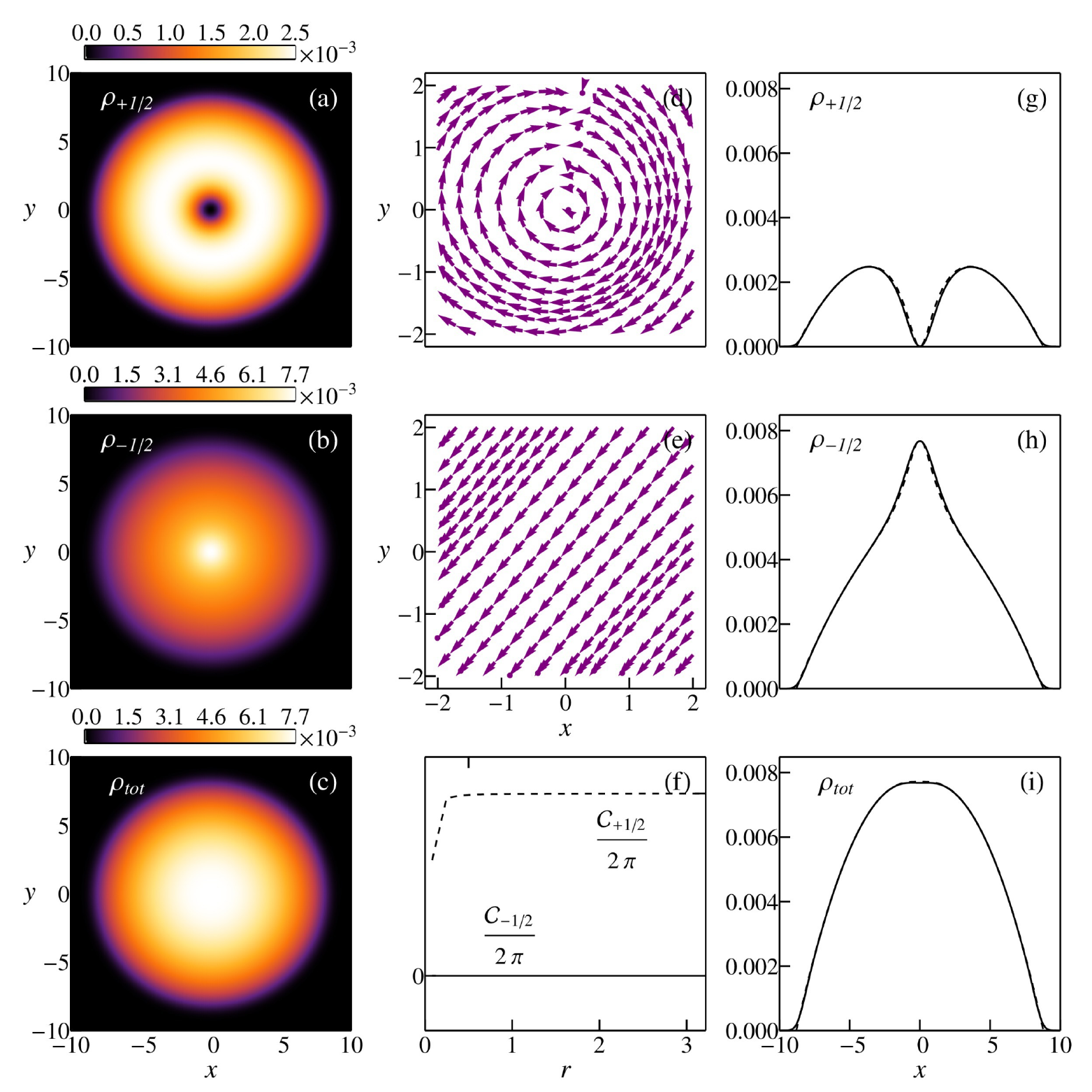}
\end{center}
\caption{(Color online) Vortex at the origin in a $j = 1/2$ BEC, generated by a dipolar $l =1$ transverse magnetic field and an axial one $B_z > 0$. The vortex of charge $Q = 1$ is in the $m = 1/2$ component. The $m = - 1/2$ component shows a density spike. The panels (a) to (f) are described in figure caption of Fig. \ref{F1}. Parameters are, $g=4000$,  $m_0 {\cal B}_1 = 1.0$ and $m_0 B_z = 1.0$. Chemical potencial is $\mu=29.85$.}
\label{F5}
\end{figure}

\begin{figure}
\begin{center}
\includegraphics[width=0.47\textwidth]{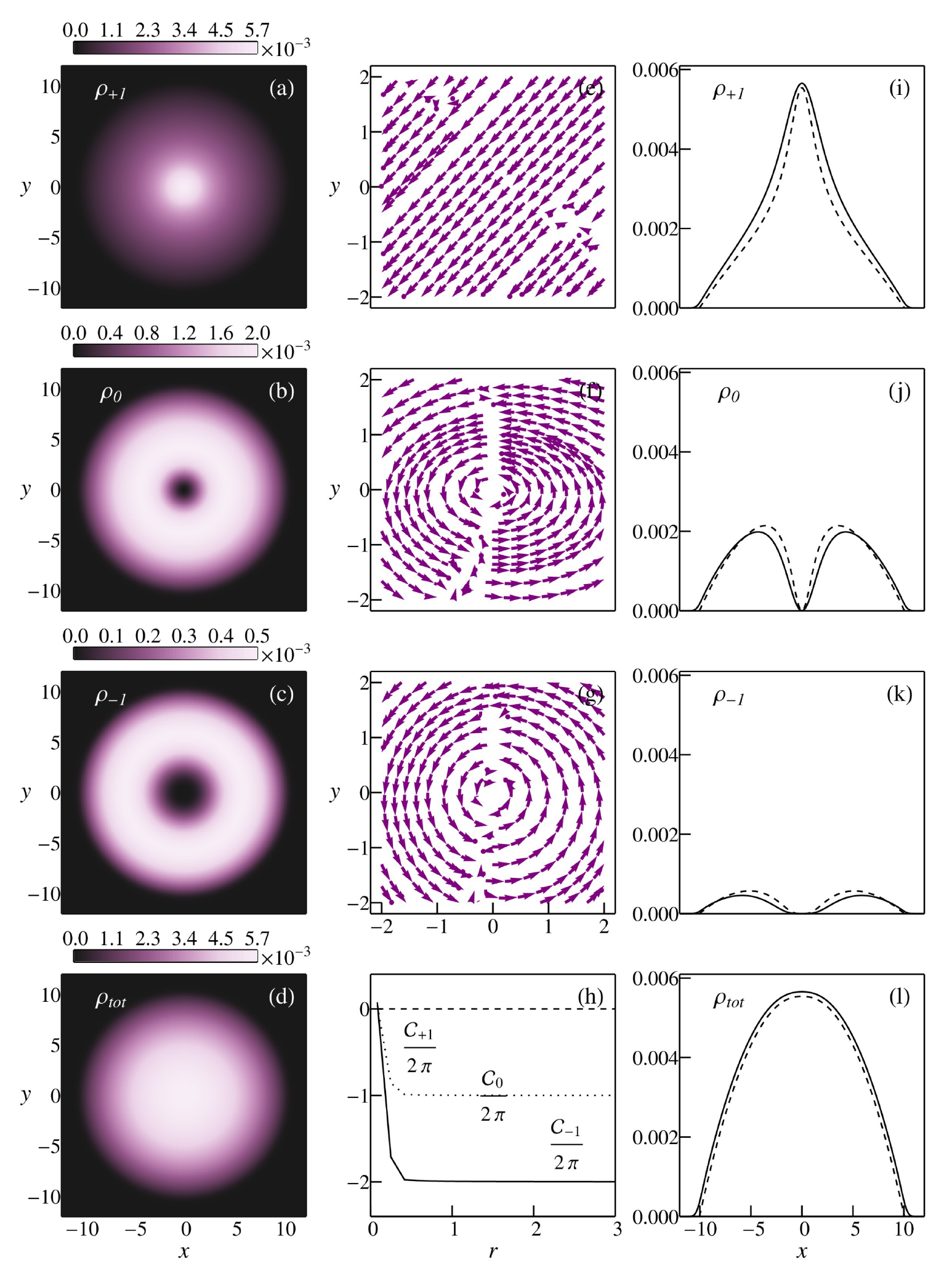}
\end{center}
\caption{(Color online) Vortices at the origin in a $j = 1$ BEC produced by a dipolar $l =1$ transverse magnetic field and an axial one $B_z < 0$. The $m = 1$ component shows a density spike, the $m = 0$ a $Q = -1$ vortex and the $m = -1$ a $Q = -2$ vortex. The arrangement of the panels is described in Fig. \ref{F3}. Parameters are, $g=8000$,  $m_0 {\cal B}_1 = 0.5$ and $m_0 B_z = -2.0$. Chemical potencial is $\mu=43.33$.}
\label{F6}
\end{figure}

\subsection{Single vortices, $B_z \ne 0$.}

As discussed in Section II, an axial field $B_z \ne 0$ breaks the symmetry and forces the appearance of the vortices in definite spin components. Moreover, the size and shape of the vortices are now dictated by the TF solution. In this subsection we show two cases with a dipolar $l = 1$ transverse magnetic field and for both $j = 1/2$ and $j = 1$ multicomponent BEC's. We show here only one case, either $B_z > 0$ or $B_z < 0$, but we have checked that the expected inverted solution indeed appears by changing the sign of $B_z$.
Figure \ref{F5} deals with $j = 1/2$. The arrangement of the panels is the same as in Fig. \ref{F1} above. Since $B_z >0$, the vortex always appears in the $m = +1/2$ component, see panels (a)-(d). However, as shown in panels (g)-(h)-(i) the density solution follows quite closely the TF solution even at the location of the vortex, in clear distinction to the case $B_z = 0$ above. As predicted in Section II, the vortex size is determined by TF solution, see Eq. (\ref{j12bz}). For the same reason, the populations $N_{\pm 1/2}/N$ are also the same as those predicted by TF as well. 

Figure \ref{F6} show the spin case $j = 1$. The description of the panels is the same as in Fig. \ref{F3}. Since $B_z < 0$, there is only one vortex configuration, namely, $Q = - 2$ at $m = -1$, $Q = 1$ at $m = 0$ and $Q = 0$ at $m = 1$, as discussed in Section II. The agreement of actual calculations, solid lines, with TF solution, dotted lines, is again quite acceptable.

\begin{figure}
\begin{center}
\includegraphics[width=0.47\textwidth]{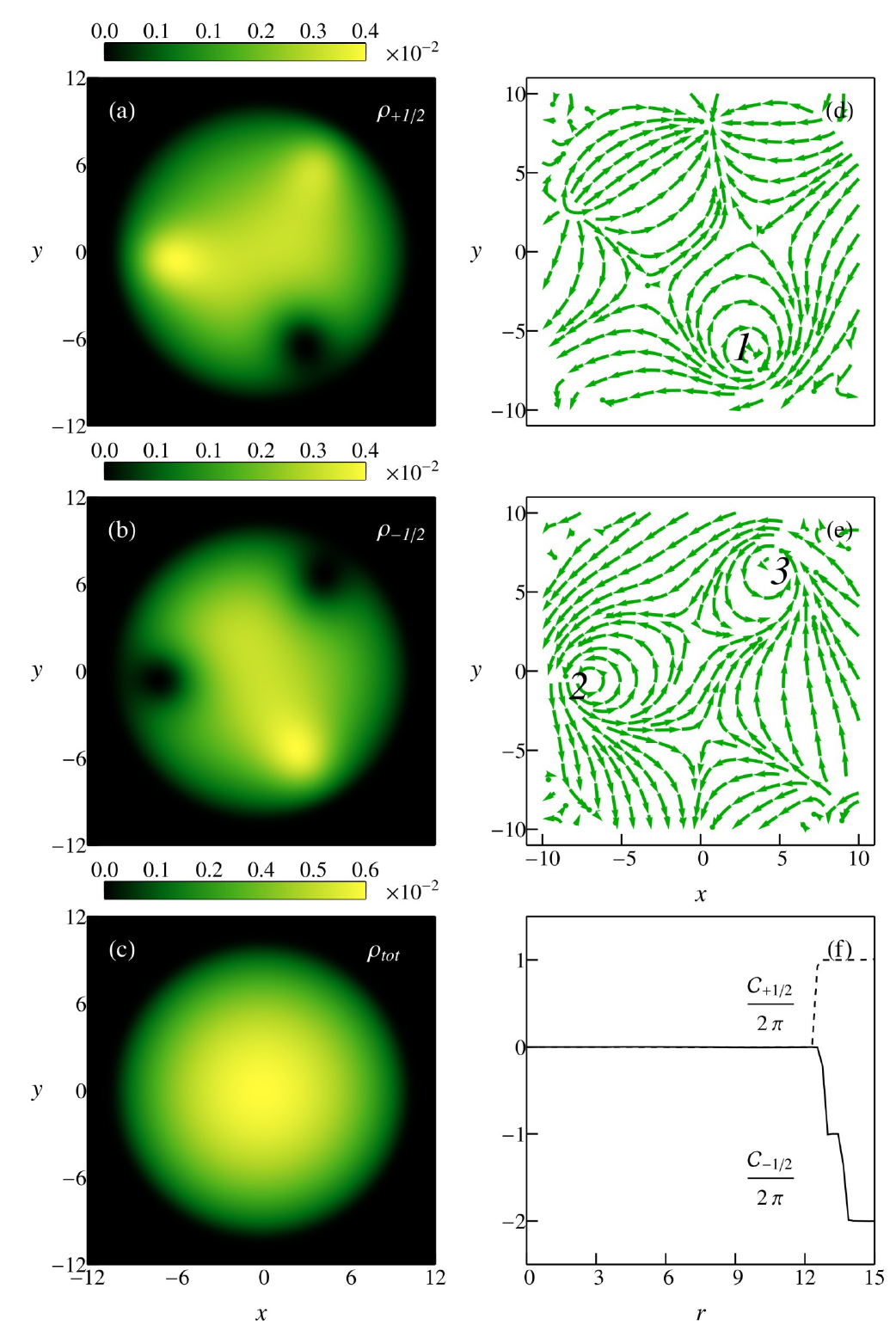}
\end{center}
\caption{(Color online) Three vortices in a $j = 1/2$ BEC, created by a transverse magnetic field with dipolar $l = 1$ zeroes at the chosen locations $\vec r_1 = (3.6,-6.1)$, $\vec r_2 = (-6.6,0.0)$ and $\vec r_3 = (3.6,6.1)$. The axial field is zero, $B_z = 0$. See text for details of the production of this field. One vortex of charge $Q = 1$ appears in the $m = 1/2$, while two vortices of charge $Q = -1$ each are present in the $m = -1/2$ component. (a) and (b) are component density plots $\rho_{\pm 1/2}$ and (c) the total one.  (d) and (e) are velocity fields $\vec v_{m}$ with the vortices locations indicated. (f) shows circulation calculations; the larger contours enclose the whole BEC cloud. Parameters are, $g = 9000$ and $m_0 {\cal B}_1 = 2.0$.  Chemical potencial is $\mu=53.41$.}
\label{F7}
\end{figure}

\begin{figure}
\begin{center}
\includegraphics[width=0.47\textwidth]{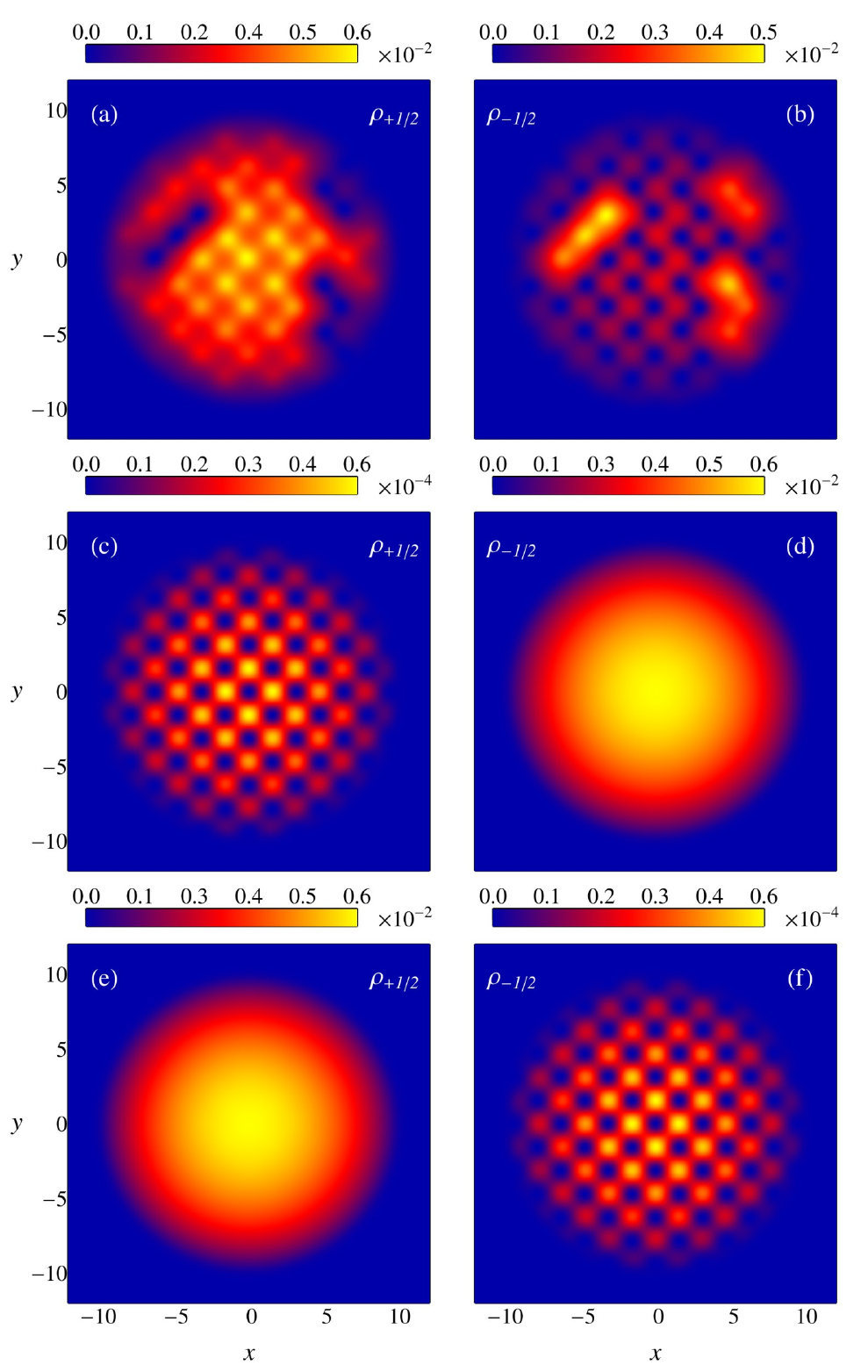}
\end{center}
\caption{(Color online) Vortex lattices in a  $j=1/2$ BEC created by the lattice transverse magnetic field given in Eq. (\ref{lattice}). Such a field has dipolar $l = 1$ zeroes in periodic sites that nucleate vortices of charge $Q = \pm 1$. We show component density plots $\rho_{\pm1/2}$. In panels (a) and (b) the axial field is zero, $B_z = 0$. In (c) and (d), the axial field is positive $B_z >0$, while in (e) and (f) negative, $B_z < 0$. Parameters are $g=8000$ $m_0 {\cal B}=2.0$ and $m_0 |B_z| = 5.0$. Chemical potentials are, respectively, 
$\mu=50.11$, $\mu=40.39$  and $\mu=40.39$. }
\label{F8}
\end{figure}

\subsection{Multiple vortices}

It was reported in the previous subsection how external fields of the form given by Eqs.(\ref{B1}) and (\ref{B2}) nucleate single vortices at the desired locations. We now turn to the possibility of nucleating arbitrary arrays of vortices.  

\begin{figure}
\begin{center}
\includegraphics[width=0.40\textwidth]{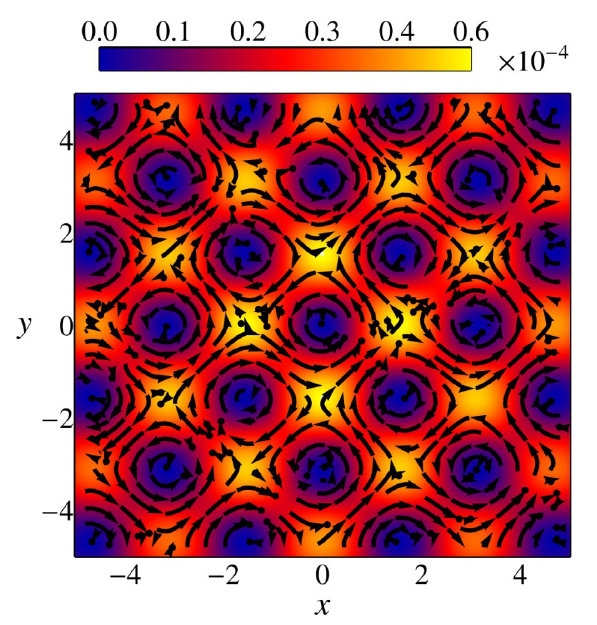}
\end{center}
\caption{(Color online) Vortex lattice corresponding to panel (c) in  Fig. \ref{F8}. The circulation field is shown in black arrows. One can observe the alternating character of the vortex topological charge. See text for details.}
\label{F9}
\end{figure}

The first situation is shown in Fig. \ref{F7}, where we study a $j =1/2$ BEC in the presence of a tailored transverse magnetic field but with no axial field, $B_z = 0$. The external transverse magnetic field is a superposition of fields generated by eight ``infinite" wires, with currents of different magnitudes and signs as indicated below, placed {\it outside} the BEC cloud  such that $\nabla \cdot \vec B = 0$ and $\nabla \times \vec B = 0$ inside the cloud region. The field of such a configuration of wires can be written as
\begin{equation}
\vec B(x,y) = \sum_{n=1}^8 I_n \frac{(x-x_n) \hat{e}_y - (y - y_n)  \hat{e}_x}{(x-x_n)^2 +(y - y_n)^2} \label{wire}
\end{equation}
where $I_n$ is the current in dimensionless units, and $\hat e_\alpha$ is the unitary vector in direction $\alpha = x, y$. For the particular configuration of Fig. \ref{F9}, the positions of the wires are on a square of size $2L$, that is, $(x_1,y_1)=(-x_3,y_3)=(L,0)$; $(x_2,y_2)=(x_4,-y_4) =(0,L)$; $(x_5,y_5)=(-x_6,y_6)=(-x_7,-y_7)=(x_8,-y_8)=(L,L)$. For the wires to be outside the BEC cloud, one must choose $L > R_{TF}$, namely, their distance from the origin is larger than the Thomas-Fermi radius $R_{TF}$. In the particular situation of Fig. \ref{F7}, $L = 16.0$ and $R_{TF} \approx 10.0$. To obtain three zeroes of the superposition of the fields (\ref{wire}) we chose the intensities of the currents as, $I_1=I_2=I_3=I_4= -2$, $I_5=I_6=1.0$ and $I_7=I_8=1.5$. The three zeroes are of dipolar order $l = 1$ and are located at $\vec r_1 = (3.6,-6.1)$, $\vec r_2 = (-6.6,0.0)$ and $(3.6,6.1)$. 

As observed in Fig. \ref{F7} there appear three vortices at the chosen zeroes of the field (\ref{wire}). In panels (a)-(b)-(c) we show density plots, while (d)-(e) are velocity fields and (f) circulation calculations. Since the axial $B_z$ field is zero, there is an spontaneous symmetry breaking and, in this particular case, there appear one vortex of charge $Q = +1$ at the component $m = 1/2$, and two vortices each of charge $Q = -1$ at the component $m = -1/2$. The circulation calculation shown in panel (f) was obtained with square contours initiating at the center until enclosing all the cloud. Thus, the circulation is ${\cal C}_{1/2}/2 \pi = +1$ at the $m= 1/2$ component, while it yields ${\cal C}_{-1/2}/2 \pi = -2$ at the $m= -1/2$ one. We have verified that in the presence of an axial field $B_z \ne 0$, all the vortices are nucleated at the same spin component. This is exemplified below with a different tailored field.

Fig. \ref{F8} shows a regular lattice of vortices created in a $j = 1/2$ BEC. We study both cases $B_z = 0$ and $B_z \ne 0$. The lattice is generated by a transverse magnetic field of the form,
\begin{eqnarray}
&&\vec B(x,y) = {\cal B} \left[ \sin({\pi x}/{ \lambda}) \cos({\pi y}/{ \lambda}) \hat{e}_x\right.\nonumber\\ 
&&\phantom{\vec B(x,y) = }\left.-\sin({\pi y}/{ \lambda}) \cos({\pi x}/{ \lambda})\hat{e}_y\right]. \label{lattice}
\end{eqnarray}
The field vanishes linearly at the points $(x,y)=(\pm n \lambda, \pm k \lambda)$ with $n$ and $k$ both integers or both half integers; that is, those points are ``zeroes" of dipolar order $l = 1$. In panels (a)-(b) there is a vanishing axial field, $B_z = 0$, and again, the spontaneous symmetry breaking mechanism nucleates arbitrary number of vortices in each component. In panels (c)-(d), the axial field is positive $B_z > 0$, while in (e)-(f) negative $B_z < 0$. Thus, respectively, the lattice is nucleated in the $m = 1/2$ or $m = -1/2$ components.

An interesting feature of the vortex lattices is shown in Fig. \ref{F9}, where we exhibit the case of Fig. \ref{F8} (c) which corresponds to vortices in the $m = 1/2$ component. In this figure we see that vortices of both charges $Q = \pm 1$ are generated within the same spin component: in the zeroes of the magnetic field that correspond to both $k$ and $n$ integers, vortices are of charge $Q=+1$, while the zeroes corresponding to both $k$ and $n$ half integers yield vortices of charge $Q=-1$. The signs of the charges are reversed if the lattice is formed in the $m = -1/2$ spin component. This behavior can be readily understood using the tools from Section II. We note that the behavior of the magnetic term in Eq.(\ref{GP-gral}) depends on the nature of the zeroes. This can be seen by expanding the magnetic field around each of its zeroes; it is of the form,
\begin{eqnarray}
	A_{kn}^{\pm}m_{0}{\cal B}_1 r \left(
	\begin{array}{cc}
	0 & e^{\pm i\phi} \\
	e^{\mp i\phi} & 0
	\end{array} 
	\right)
\end{eqnarray}
where $A_{kn}^{\pm}$ is a constant. The upper sign case corresponds to both $n$ and $k$ integers, while the lower one to the half integer one, and thus, the sign of the charge depends not only on the spin component but on the values of $k$ and $n$. An ``undesired" characteristic of this lattice is that $\nabla \times \vec B$ is not zero in the whole region occupied by the BEC cloud. That is, one needs ``sources" in the region occupied by the condensate. Nevertheless, this case illustrates the possibility of generating vortices at the desired locations. We leave the tailoring of the corresponding fields to the ingenuity of the experimental researchers. 

\section{Final Remarks}

As we have seen, in multicomponent BEC clouds, the presence of transverse magnetic fields with vanishing values at given spatial locations, nucleate vortices at those very points. The vortices can thus be placed at any location ``on demand". For single vortices, one can simply choose the zero of the fields, such as those in Eqs. (\ref{B11}) and (\ref{B12}), at any location and a vortex will be nucleated there. For multiple vortices, one can set multiple wires, such as in Eq. (\ref{wire}), and by changing the current intensities almost at random, one finds different zeroes that, again, nucleate vortices. The same comment applies to any lattice with zeroes of the field. In addition, we have verified the further prediction that the total density $\rho(x,y) = \left| \psi_+(x,y) \right|^2 + \left| \psi_-(x,y) \right|^2$ does not show any evidence of the vortices and that its form is the ``simple" TF solution of GP in a harmonic potential.

It is interesting to contrast the present form of the appearance of the vortices with that of a rotating superfluid. In the rotating frame, the (one-component) GP equation looks like\cite{castin},
\begin{equation}
 -\frac{\hbar^2}{2m} \nabla^2 \Phi(\vec r) +  (\vec \Omega \cdot \vec L )\Phi(\vec r) + g \> \left| \Phi(\vec r) \right|^2 \Phi(\vec r) - \mu \Phi(\vec r) = 0  ,\label{GP-8}
\end{equation}
where $\vec \Omega = \Omega \hat z$ is the angular velocity of rotation and  $\vec L$ is the angular momentum operator, 
\begin{equation}
\vec L = \frac{\hbar}{i} \vec r \times \nabla .
\end{equation}
By a vector identity one can rewrite,
\begin{equation}
(\vec \Omega \cdot \vec L) \Phi(\vec r) = \frac{\hbar}{i} \left(\vec \Omega \times \vec r\right) \cdot \nabla \Phi(\vec r) 
\end{equation}
and, therefore, one can see that the rotation term is equivalent to introducing a vector potential $\vec A = \vec \Omega \times \vec r$ corresponding to an axial uniform ``magnetic field" $\vec B = \vec\Omega$ (in appropriate units). 

In its celebrated paper, Abrikosov\cite{abrikosov} showed that if the above equations hold, a lattice of vortices, of the type described in Section I,  will be created by the presence of such a magnetic  or appropriate gauge field. It does appear, therefore, that vortices are created by the coupling of the {\it orbital} angular momentum $\vec L$ with the external field $\Omega$, i.e. $\vec \Omega \cdot \vec L$. Since many vortices are created, and the global state of the gas must be the most stable one, the developed vortices then arrange themselves in a triangular lattice due to the repulsive
interaction among them\cite{abrikosov2}.

In the case described in this work, vortices also appear because of a magnetic angular momentum coupled to an external gauge field. It does appear, however, that in the present case it is the {\it intrinsic} or {\it spin} angular momentum the main player instead of the orbital angular momentum. 

We believe that the above two cases may not be as different as they may appear. This assertion may be based on the observation that GP 
equation near a zero of the magnetic field can be transformed as one with an artificial gauge field coupled to the orbital angular momentum. Let us see this briefly. For simplicity, take the case $j=1/2$, $B_z = 0$, with $\kappa = m_0 {\cal B}_1$. The coupled equations (\ref{GP-gral}) may be written as
\begin{equation}
\left[ -\frac{\hbar^2}{2m} \nabla^2 + V_{ext}(\vec r) + g |\Psi|^2  + \kappa \left(x \sigma_x + y \sigma_y\right) -\mu\right] \Psi =0. \label{LG1}
\end{equation}
where
, 
\begin{equation}
\Psi=\left(
\begin{array}{c} \psi_{+}  \\ \psi_{-} \\
\end{array} 
\right) ,
\end{equation}
and
$|\Psi|^2=|\psi_+|^2+|\psi_-|^2$.
By applying the unitary transformation
\begin{equation}
U = e^{-i \phi \sigma_z/2} e^{-i \pi \sigma_y /4} e^{i \phi \sigma_z/2} ,
\end{equation}
with $\tan \phi = y/x$, equation (\ref{LG1}) now becomes\cite{longuett,VRR89}
\begin{eqnarray}
\left[ \frac{1}{2m} (-i\hbar \nabla - \vec A_x \sigma _x - \vec A_y \sigma_y -\vec A_z \sigma_z)^2 \right.
\nonumber\\
\left.+ V_{ext}(\vec r) + g|\tilde \Psi|^2 + \kappa r \sigma_z -\mu\right] 
\tilde\Psi =0 ,\label{LG}
\end{eqnarray}
with $\tilde \Psi$ the transformed state. The induced artificial gauge potentials may be easily calculated\cite{VRR89}. In particular, the diagonal potential $\vec A_z$ is,
\begin{equation}
\vec A_z = - \frac{\hbar}{2r}\hat e_\phi
\end{equation}
which corresponds to a nonuniform {\it singular}  magnetic field, actually, to that of a magnetic monopole. 
Thus, the vortex nucleation in the rotating condensate appears to be equivalent to ours via an artificial magnetic monopole field. This, in turn, is intimately related to Berry phases\cite{berry,larson}. Therefore, the present scheme may also be seen from the perspective of artificial gauge fields\cite{spielman,dalibard,juzeliunas2,murray,spielman2}, although it should be clear that we do not assume adiabaticity of the diagonal terms in (\ref{LG}).

Because of the scope of the article, we did not present results we have found in 3D BEC's. If the magnetic fields are chosen just as in the 2D case, on finds that  the vortices are of cylindrical nature in the $z-$direction. This is because the nucleation points (where the transverse magnetic field becomes zero) are actually filaments that cross the full BEC cloud.

 We believe the present results can be simply realized in actual experimental conditions. In particular, for a $^{87}$Rb BEC there are two cases that can be simply matched. One is the $F = 1$ case that naturally corresponds to the $j =1$ situation here discussed. The other, the case $F = 2$ may be tailored to have only the components $m_1 = 1$ and $m_2 = 2$ corresponding to $j = 1/2$. The external magnetic fields may be produced by single-wire magnetic fields, as given by (\ref{wire}). However, an additional non-uniform small magnetic field in the $z$ direction should be used to separate the spin components such that the phenomena may be observed.

The present scheme suggests also an alternative route to quantum turbulence. Recently, Bagnato et al.\cite{vanderlei} found a way to stir a $^{87}$Rb condensate by means of time dependent magnetic fields, achieving a state of quantum turbulence. Although we are not asserting that our procedure explains those experimental results, it does appear that once vortices are nucleated by transverse magnetic fields, the vortices may be stirred by simply changing in time the positions of the zeroes of the magnetic field. It seems evident that a strong stirring of the vortices may lead to a quantum turbulent state. The study is under way and will be reported elsewhere. 

We thank support from grant DGAPA UNAM IN108812. R.Z.Z. and M.L.H acknowledge support from CONACYT, Mexico.

\end{document}